\documentclass[%
 aip, rsi
 amsmath,amssymb,
 reprint, floatfix
]{revtex4-1}

\usepackage{graphicx}% Include figure files
\usepackage{dcolumn}% Align table columns on decimal point
\usepackage{bm}% bold math
\usepackage[utf8]{inputenc}
\usepackage[T1]{fontenc}
\usepackage{mathptmx}
\usepackage{braket}
\usepackage{float}
\usepackage{tasks}
\usepackage{xcolor}
\usepackage{dblfloatfix}
\usepackage{array}
\usepackage{makecell}

\begin{document}

\preprint{AIP/123-QED}

\title{Observing sub-Poissonian statistics of twisted single photons using oscilloscope}
% Force line breaks with \\

\author{Nijil Lal}
\email{nijil@prl.res.in}
\affiliation{Physical Research Laboratory, Ahmedabad 380009, India}
\affiliation{Indian Institute of Technology, Gandhinagar 382355, India}

\author{Biveen Shajilal}
\altaffiliation[Present address: ]{Centre for Quantum Computation and Communication Technology, Research School of Physics and Engineering, The Australian National University, Canberra, Australian Capital Territory 2601, Australia}
\affiliation{Cochin University of Science and Technology, Kochi 682022, India}

\author{ Ali Anwar}
\altaffiliation[Present address: ]{Centre for Quantum Technologies, National University of Singapore, 3 Science Drive 2, S117543, Singapore}
\affiliation{Physical Research Laboratory, Ahmedabad 380009, India}
\author{ Chithrabhanu Perumangatt}
\altaffiliation[Present address: ]{Centre for Quantum Technologies, National University of Singapore, 3 Science Drive 2, S117543, Singapore}
\affiliation{Physical Research Laboratory, Ahmedabad 380009, India}

\author{R. P. Singh}
\affiliation{Physical Research Laboratory, Ahmedabad 380009, India}

\date{\today}% It is always \today, today,

\begin{abstract}
Heralded single photon sources (HSPS) from spontaneous parametric down-conversion are widely used as single photon sources. We study the photon number statistics of an HSPS carrying orbital angular momentum in our laboratory and observe the sub-Poissonian statistics using only photo detectors and an oscilloscope. 
\end{abstract}

\maketitle

\section{\label{sec:level1}Introduction}

Photon is a prominent qubit candidate in quantum information processing (QIP). Most of these applications require a single photon source (SPS) which will provide an ``on-demand", deterministic supply of photons. Such ideal SPSs can be realized in crystal color centers \cite{Wratchtrup}, quantum dots \cite{qd} single atoms \cite{hennrich}, single ions \cite{maurer}, and single molecules \cite{steiner}. Photon being an indivisible quantum of energy and the consequent impossibility of copying associated information from an unknown arbitrary state makes single photon sources an ideal choice for quantum information processing, secure communication, and metrology. Spontaneous parametric down-conversion (SPDC) is one of the most widely used processes to generate single photon sources as it provides a robust and bright source at room temperature \cite{fasel2004high}. In this process, a pump photon is down-converted in frequency in a $\chi^{(2)}$ nonlinear crystal to a pair of photons, conventionally called as signal (\textit{s}) and idler (\textit{i}). The detection of one photon from the pair heralds the presence of another, and this conditioned detection serves as a technique for generating single photon sources. SPDC is also important in QIP as a bright source of entangled photon pairs \cite{jabir} and thus finds applications in quantum teleportation \cite{boschi}, quantum cryptography \cite{gisin}, and quantum dense coding \cite{dense}.

A single photon source shows non-classical properties such as sub-Poissonian photon number distribution and anti-bunching. The inherent thermal nature of the photon distribution in the individual arms (i.e., signal or idler) evokes the possibility of occurrences of multiple photons and hence would qualify the source as less quantum than it is expected to be. The occurrence of multiple photon pairs and consequent photon bunching threatens the security in cryptography related applications. Thus, it is important to study and quantify the non-classicality of the single photon sources being used. A general classification puts the sources into three categories, super-Poissonian, Poissonian and sub-Poissonian according to the photon number distribution. Classical sources, like thermal and coherent radiation, show super-Poissonian and Poissonian statistics respectively while a non-classical source, such as a single photon source, shows sub-Poissonian distribution. Violations of bounds defined for the number statistics of classical sources imply the non-classical behavior of single photon sources \cite{shortmandel,rarity,laurat,perina}. Photon antibunching is another manifestation of the non-classical behavior of the source and can be observed through the measurement of the degree of second-order coherence, $g^{(2)}{(\tau)}$. Single photon sources show antibunching and hence $g^{(2)}{(0)}$ falls to zero. Though antibunching and sub-Poissonian statistics are completely non-classical effects and tend to occur together in many systems, they are distinct effects and need not necessarily be associated with one another \cite{singh,zou}. For a heralded SPS, antibunching can be observed using a heralded Hanbury Brown - Twiss (HBT) type experiment \cite{grangier,u'ren}. Twisted single photons carrying orbital angular momentum (OAM) is gaining interest in quantum information processing as they provide infinite number of orthogonal states to encode information \cite{mirhosseini2015,barreiro}. The statistical second order correlation of classical beams \cite{ashokcorr,ashokhbt} as well as photons \cite{Lal} carrying OAM has already been studied. In the present work, we verify the sub-Poissonian nature of heralded single photons generated in parametric down-conversion having Gaussian as well as higher order OAM using a simple setup consisting of photodetectors and an oscilloscope.

\section{Theory}
\label{S:2}

One can define an ideal single photon source as a Fock state for which the number statistics gives mean, $\mu =1$ and variance $\sigma^2=0$. Treatment of the electromagnetic field as a quantum harmonic oscillator allows us to define quadrature field operators, $\hat{X}_{1}$ and $\hat{X}_{2}$, which can be shown to follow the uncertainty relation \cite{gerry}, $ \braket{\Delta \hat{X}_{1}^2} \braket{\Delta \hat{X}_{2}^2} \geq 1/16 $ . The minimum uncertainty state, with $\braket{\Delta \hat{X}_{1}^2}=\braket{\Delta \hat{X}_{2}^2}$, is the coherent state,

\begin{equation}
\ket{\alpha}=\exp^{\frac{{-|\alpha|}^2}{2}}  \sum_{n=0}^{\infty}\frac{\alpha ^n}{\sqrt{n!}}\ket{n}
\end{equation}
which defines a Poissonian distribution of photon number \cite{glauber}, $n$, with $\langle \left(\Delta \hat{n}\right) ^2 \rangle= \langle \hat{n} \rangle $. Squeezing these states in terms of number or phase is achieved when the respective variance is less than that corresponding to the coherent state. The photon number variance can be written as,
\begin{equation}
\left\langle{\Delta \hat{n}}^2 \right\rangle= \langle \hat{n} \rangle + \langle \hat{a}^\dagger \hat{a}^\dagger \hat{a} \hat{a} - \langle \hat{a}^\dagger \hat{a} \rangle ^2 \rangle
\end{equation}
where $\hat{a}$ and $\hat{a}^\dagger$ are the annihilation and creation operators.
Based on the variance, the Mandel Q-parameter \cite{mandelbook} is defined by,
\begin{equation}
Q \equiv \frac{\langle \hat{a}^\dagger \hat{a}^\dagger \hat{a} \hat{a} - \langle \hat{a}^\dagger \hat{a} \rangle ^2 \rangle}{\langle \hat{a}^\dagger \hat{a} \rangle} = \frac{\langle \left(\Delta \hat{n}\right) ^2 \rangle - \langle \hat{n} \rangle}{\langle \hat{n} \rangle}  
\end{equation}

It classifies light sources on the basis of photon number fluctuations. $Q \geq 0$ for coherent and all other classical sources of light while sub-Poissonian light is identified with $Q < 0 $ \cite{mandelbook,mandel}.
\section{Experiment}
\label{S:3}

\makeatletter
\newcommand{\mypm}{\mathbin{\mathpalette\@mypm\relax}}
\newcommand{\@mypm}[2]{\ooalign{%
  \raisebox{.1\height}{$#1+$}\cr
  \smash{\raisebox{-.6\height}{$#1-$}}\cr}}
\makeatother

\begin{figure}[h]
\begin{center}
\includegraphics[width=\linewidth]{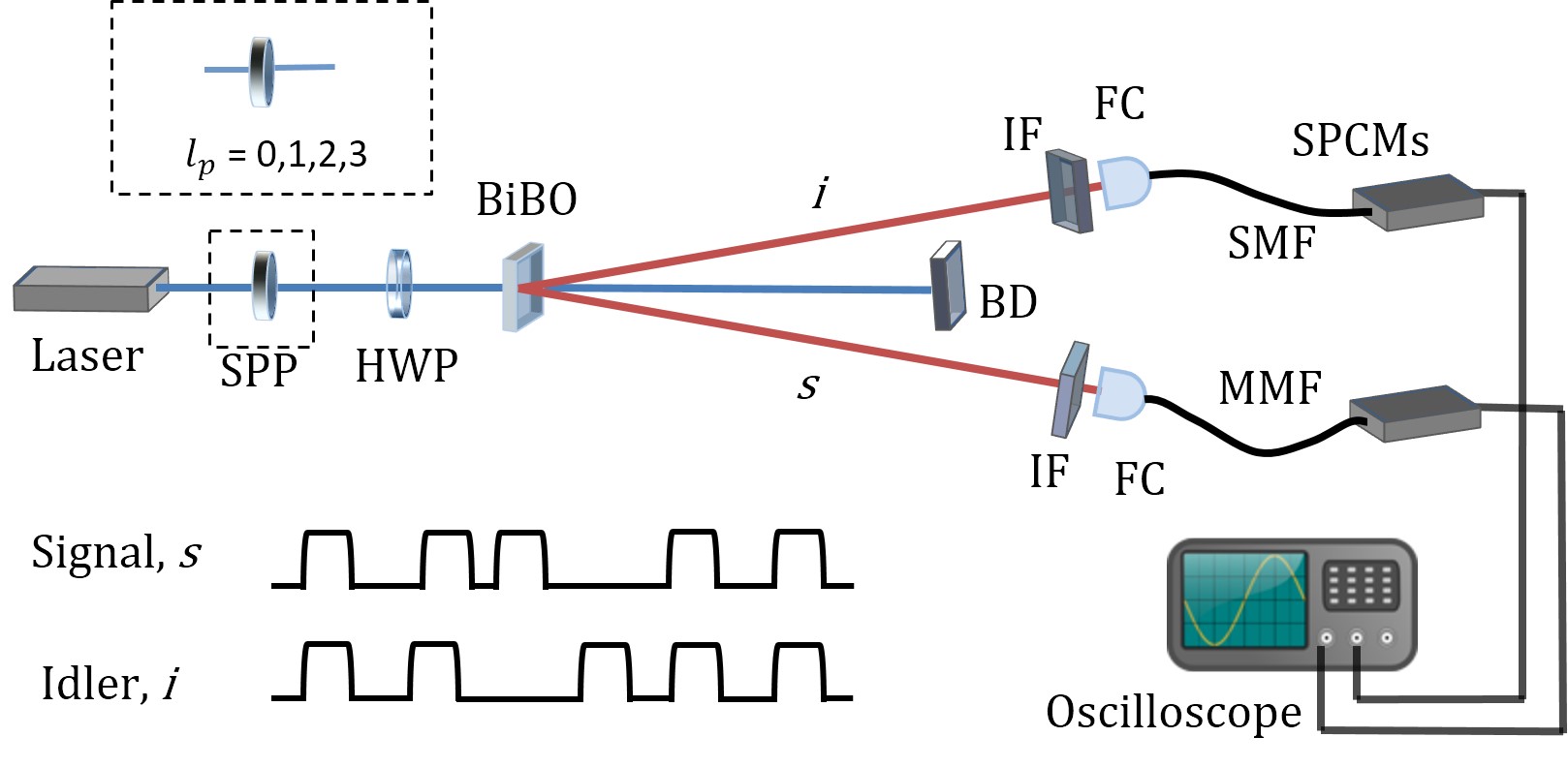}
\caption{Experimental setup to determine the photon number statistics of the heralded PDC source. HWP - half wave plate, BiBO - non linear crystal, BD - beam dump, IF - interference filters, FC - fiber couplers, SPCMs - single photon counting modules. Different OAM orders can be imparted to the heralded single photon by placing SPPs of different orders in pump and projecting idler photon in a single mode fiber.}
\label{fig:setup_final}
\end{center}
\end{figure}

The experimental setup used for building the statistics of the heralded single photon source is given in Fig.\ref{fig:setup_final} \cite{lal2016}. A 405 nm diode laser (Toptica iBeam Smart, 5 mW) is used as a pump to generate photon pairs from a Type-I BiBO crystal (Castech, 5 mm thickness). The polarization of the pump is oriented along the crystal axis using a half-wave plate. The output pairs are spectrally filtered using interference filters of passband $810\pm5$ nm to rule out the non-degenerate pairs. Two diametrically opposite regions are selected from the cone of correlated pairs of photons generated in the Type-I SPDC. These individual arms (signal and idler) are then collimated and coupled to a multimode fiber (ThorLabs, M43L02) through a fibre coupler (CFC-5X-B).

The individual signal and idler photons are detected using single photon detectors. We have used two single photon counting modules (SPCM, Excelitas SPCM-AQRH-16-FC) placed in both signal and idler arm for the detection of the photon pairs. Photomultiplier tubes (PMTs), which will be more common in laboratories than an SPCM also can be used for the detection. The coincidences are maximized by optimizing the coupling to the fiber. The SPCM (or PMT) generates a TTL pulse in response to each photon incidence. However, the total number of photon incidences that can be recorded by a detector is limited by the detector dead time (22 ns for the SPCM used).

The output waveform from each arm of the SPDC is recorded using a digital oscilloscope (Infiniium 90000A series) for an exposure time of 20.5 ms. 10 iterations of such time series have been recorded to calculate the $Q$-parameter. Each iteration records a time series of 20.5 Mpts and a resolution of 1 ns. Oscilloscopes with lower specifications can also be used but with compromising the length of the time series to be captured and time resolution. The output from the detectors, as observed through the oscilloscope, consists of TTL pulses corresponding to photon incidences. Since the photon counters are not photon number-resolving, the output pulses can correspond to the incidence of one or more photons. The single photon detectors/PMTs produce an electrical signal in response to the absorption of light energy. It is important to choose intensity of the source and measurement window such that the detector measures one photon on an average for each incidence of the signal. The pump is kept in the low power regime (pump power = 1 mW) in order to avoid such scenarios due to the possible multiple photon generations \cite{razavi}. 

\begin{figure}[h]
\begin{center}
\includegraphics[width=\linewidth]{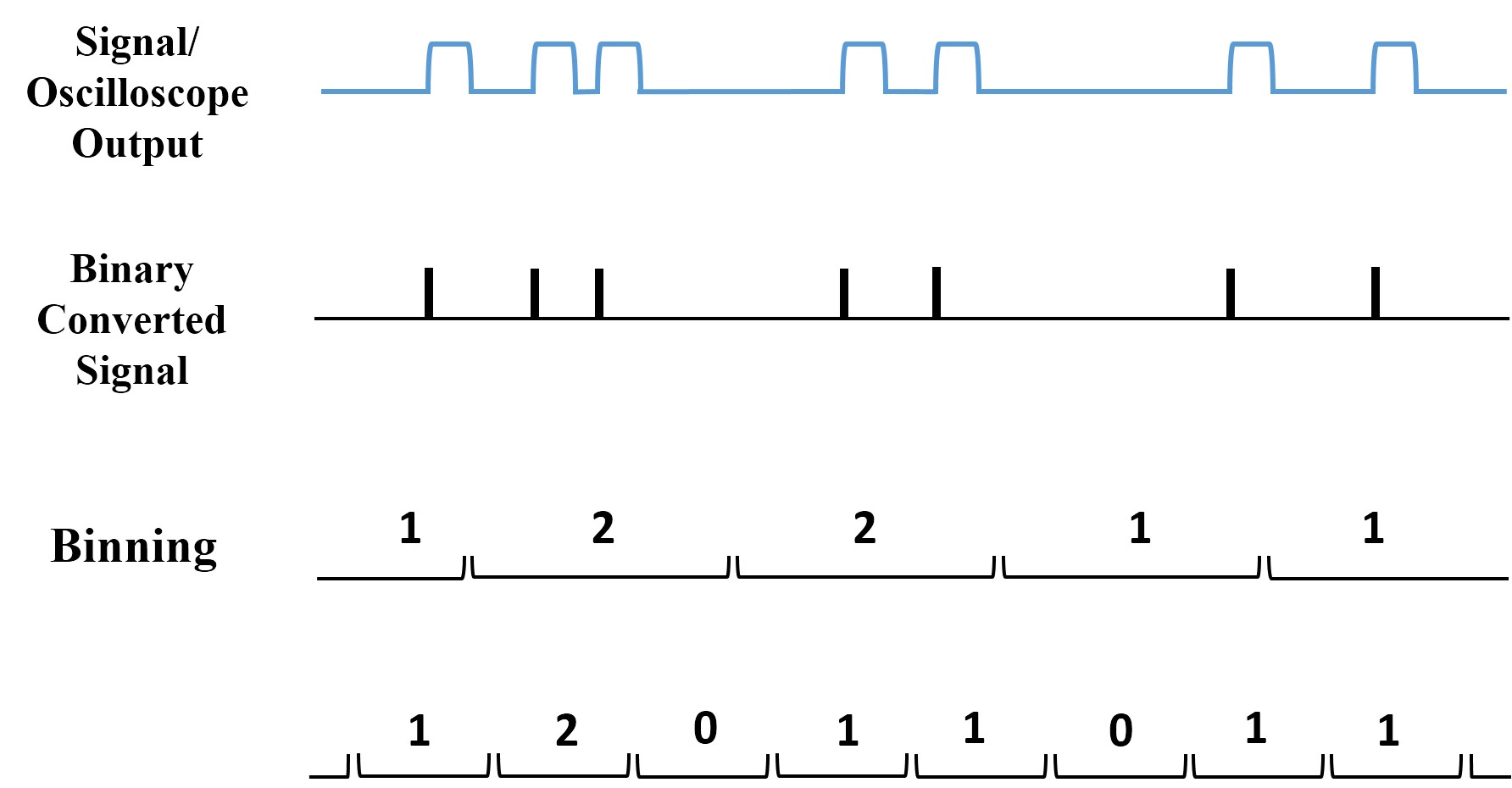}
\caption{Converting the signal to a binary time series. The average photon number is dependent on the size of the binning window.}
\label{fig:Binning}
\end{center}
\end{figure}

In order to build the statistics, the detection events  at the photo diodes are counted within a defined time interval (binning window) and the distribution of the number of such occurrences is obtained. This is done by converting the analog output from the oscilloscope into a series of 1s and 0s which corresponds to the presence or absence of a detection. We built a MATLAB code which identifies an event pulse in the oscilloscope output time series and labels its onset as 1 and the rest of the series as 0. This generates a binary time series of events in terms of the occurrences or non-occurrences of photon detection (Fig.\ref{fig:Binning}). A probability distribution of events is built from this generated time series.

The time series of binaries generated for both the signal and idler are then sliced into smaller binning windows in such a way that statistically one detection falls in a bin. The number distribution is developed for these individual arms. A new time series corresponding to the coincidences of photon pairs is developed from the time series corresponding to signal and idler, in a way that a count is observed in one arm (within the coincidence window, $\tau_c$ = 1 ns) given a detection in the other arm. The number distribution of this new time series corresponds to the statistics of the heralded single photon source from SPDC. The matlab code to build the statistics from the recorded time series is shared in GitHub repository for open access\cite{subPgithub}.

Twisted single photons of different orders are obtained in SPDC by pumping with optical vortices of different orders and projecting the idler photon in a single mode fiber and hence selecting only those with OAM, $l$ =0. Due to OAM conservation in SPDC, this will result in only those signal photons which have the same OAM as the pump contributing to the coincidence counts \cite{mair}. Hence, a photon number distribution obtained for the coincidence events will correspond to the number statistics of single photons carrying OAM.

\section{Results and Discussion}
\label{S:4}

Initially, to see the photon number distribution for a classical source, we select a coherent laser source (Thorlabs 2mW HeNe, 632.8 nm). The intensity is attenuated such that the detected photo-counts are below the saturation level of the detectors. The time series of events recorded using an oscilloscope was converted into a binary series of photon incidence events.

\begin{figure}[h]
\begin{center}
\includegraphics[width=\linewidth]{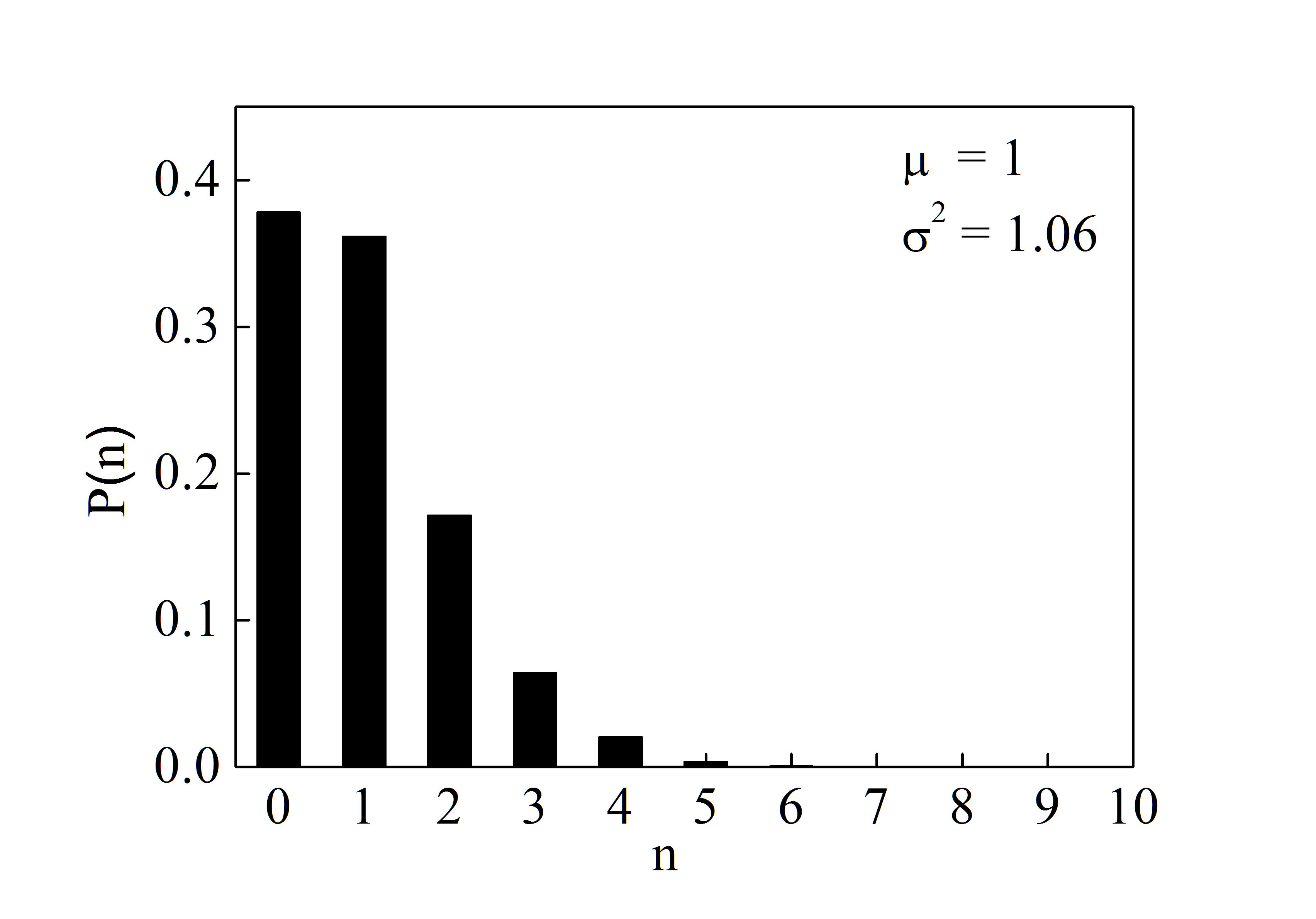}
\caption{Photon number distribution for a coherent source. Mean photon number is approximately equal to the variance.}
\label{fig:Poiss}
\end{center}
\end{figure}
A counting histogram that gives the probability of detection of \textit{n} events within the corressponding time bin is generated for this time series that shows Poissonian distribution (Fig.\ref{fig:Poiss}). For a time bin width such that an average number of 1 photon falls within the bin ($\mu = 1$), the variance ($\sigma^2$) is determined to be $1.06$. The $Q$-parameter is $0.06$ which is close to zero as expected for a Poissonian distribution of photon numbers.

\begin{figure}[h]
	\centering
	\hspace{-0.8cm}
	\begin{minipage}[h]{0.5\textwidth}
 	\centering
 	\includegraphics[width=\linewidth]{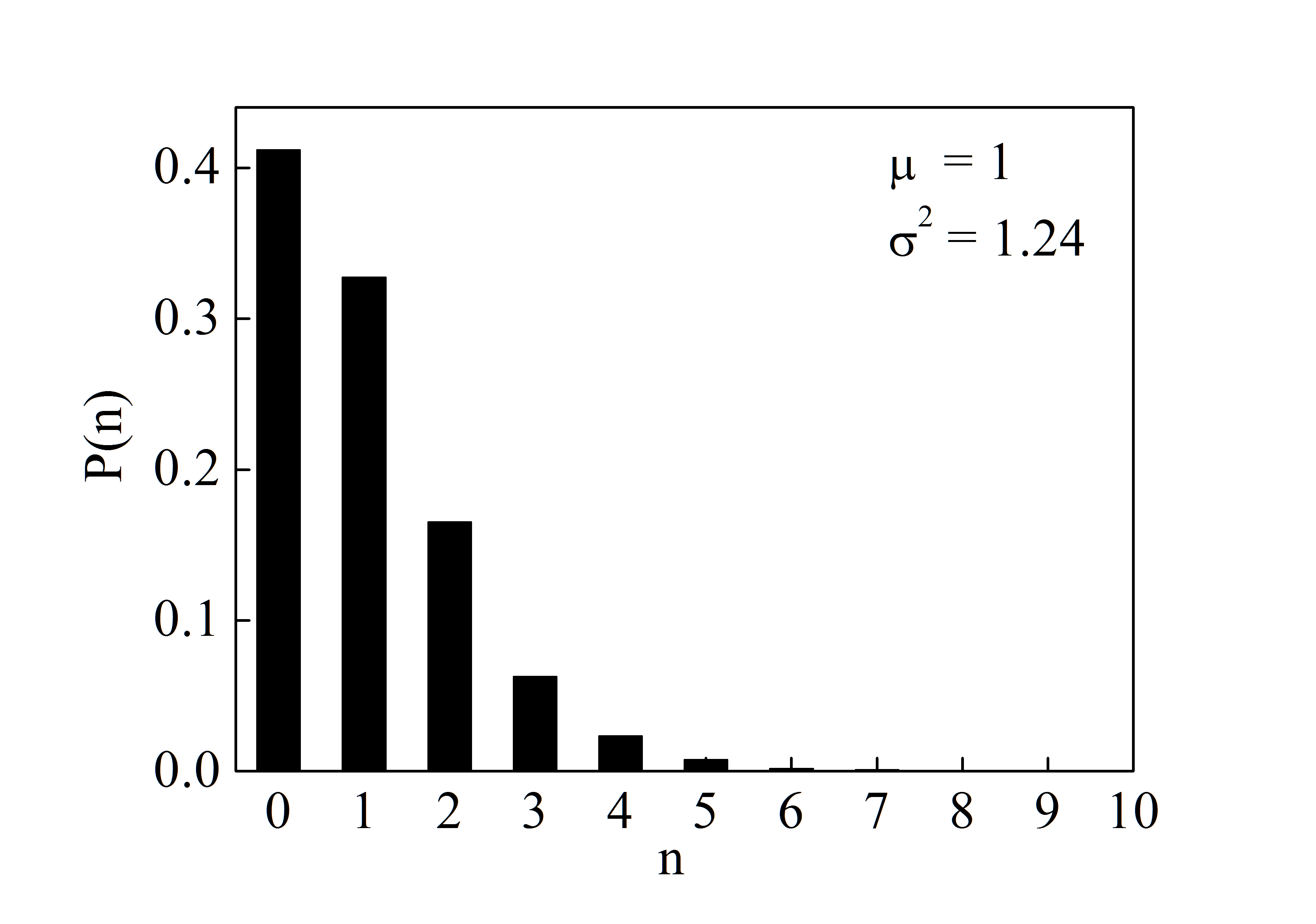}
	\label{fig:sub:super_Pid}
 	\end{minipage}
	\hspace*{\fill}
	\hspace{-0.8cm}
	\begin{minipage}[h]{0.5\textwidth}
	\centering
	\includegraphics[width=\linewidth]{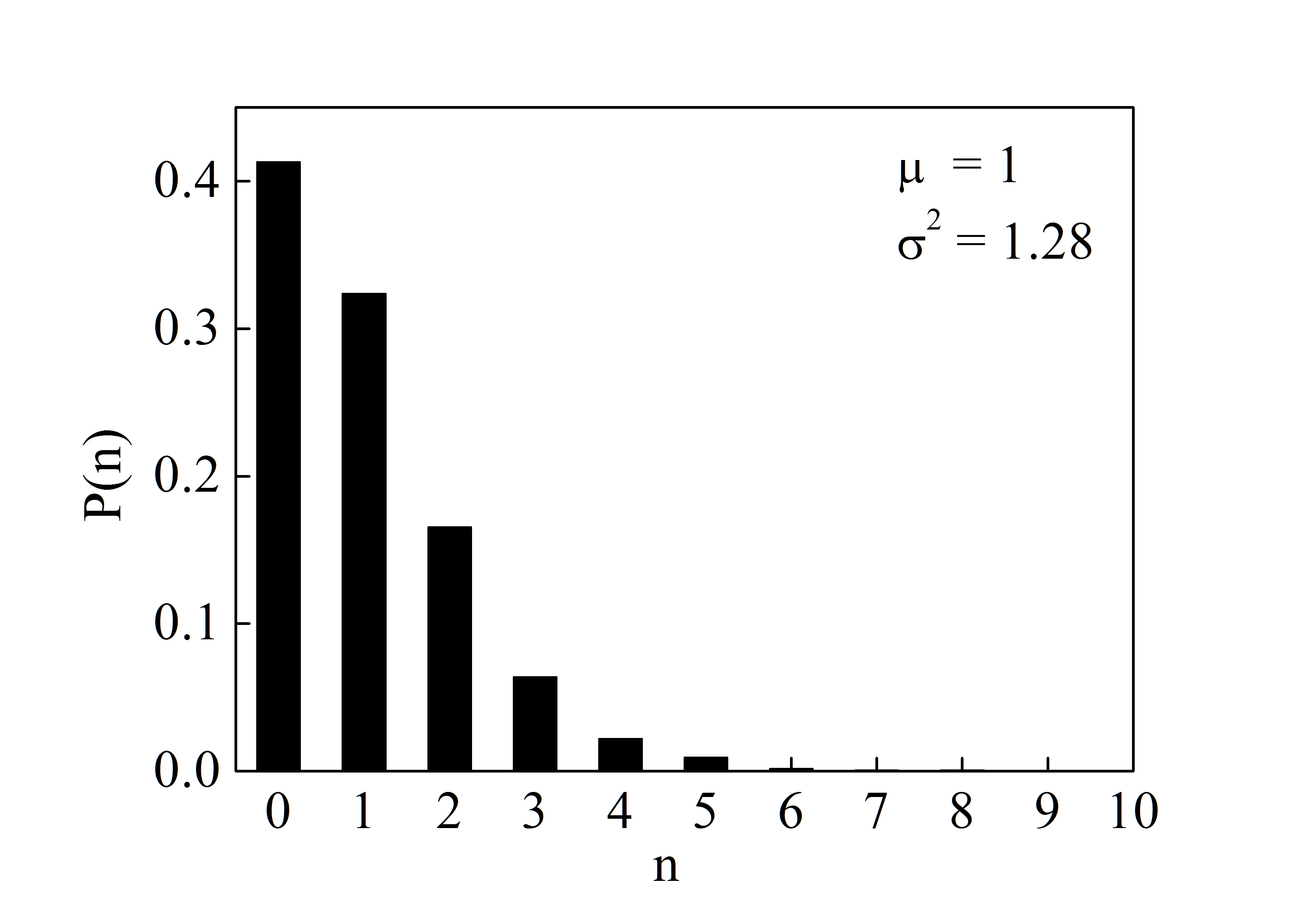}
	\label{fig:sub:super_Psg}
  	\end{minipage}
	\hspace*{\fill}

 	\caption{Photon number distribution for thermal photons in the individual idler (top) and signal (bottom) arms of the down converted pairs.}
    	\label{fig:super_P}
\end{figure}

In the same way, photon number histograms are generated for the individual arms of the down-converted pairs of photons (Fig.\ref{fig:super_P}). For an average photon number $\mu = 1$, the variances were $1.24$ and $1.28$ respectively for the signal and idler arms which are significantly apart from that corresponds to a Poissonian distribution. The $Q$-parameter is $0.24$ and $0.28$ respectively. This super-Poissonian behaviour is expected since the individual signal and idler photons from the down-conversion process are in a thermal state\cite{blauensteiner,guothermal,avenhausthermal}.

\begin{figure}
\begin{center}
\includegraphics[width=\linewidth]{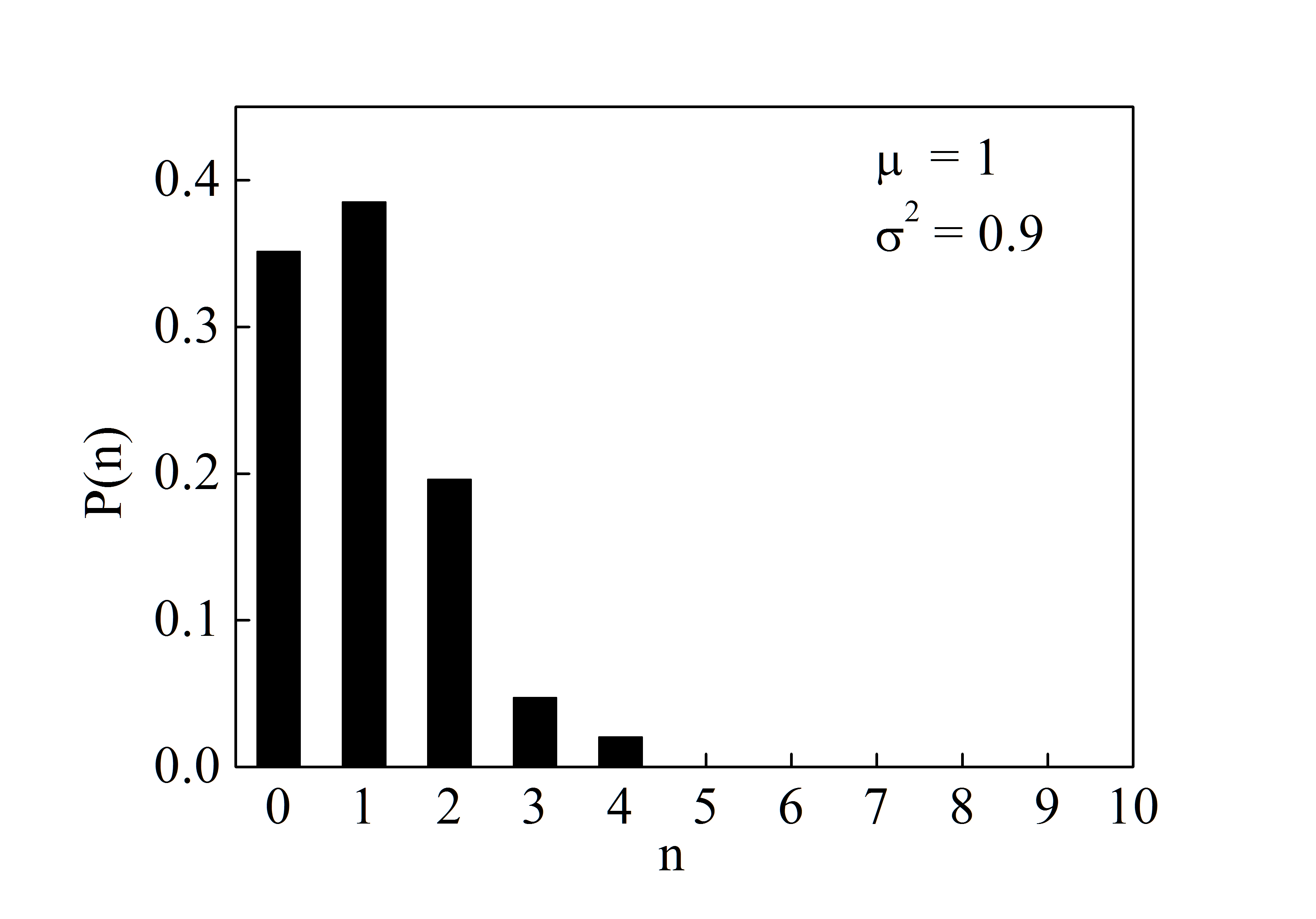}
\caption{Photon number distribution for a heralded single photon source.}
\label{fig:subP}
\end{center}
\end{figure}

For the heralded single photon source, the number statistics is built from the binary series of coincidences generated by post-processing the recorded events of individual signal and idler photons (Fig.\ref{fig:subP}). For $\mu = 1$, the variance ($\sigma^2$) for this source is calculated to be 0.9. The Mandel Q-parameter comes out to be $-0.10$. The negative value of Q is indication of the sub-Poissonian behavior of the source. A set of repeated measurements reveals the average value, $Q$ = -0.099 $\pm$ 0.004 for a heralded SPS generated using a Gaussian pump profile.

Q-parameter for the heralded single photons having different OAM, obtained by pumping with an optical vortex and projecting the idler arm to $l$ = 0, are given in Table \ref{tab:subP_vortex}. Time series of photon detection events of length 20.5 ms are recorded using the oscilloscope. 10 such iterations are recorded to determine the $Q$-parameter within statistical errors. For heralded single photons, it is found to violate the classical bounds of number statistics (i.e., $Q \geq 0$) by tens of standard deviations for photons carrying different OAM. The $Q$-parameter is observed to decrease with increasing order of twisted photon OAM. This can be attributed to the reduction in the number of coincidence events with higher order pump as the singles in the idler arm (with $l$ = 0) decreases with pump order \cite{anindya}.

\begin{table}[h]
    \centering
\begin{tabular}{ | m{3cm} | m{3cm}| } 
\hline
\thead{Twisted photon \\ OAM, \textit{l}}& \thead{Q-parameter} \\ 
\hline
\thead{$0$} & \makecell[c]{$-\hspace{0.07cm}0.099$\\ $\pm$ $0.004$} \\
\hline
\thead{$1$} & \makecell[c]{$-\hspace{0.07cm}0.101$\\ $\pm$ $0.001$} \\ \hline
\thead{$2$} & \makecell[c]{$-\hspace{0.07cm}0.103$\\ $\pm$ $0.003$} \\ \hline
\thead{$3$} & \makecell[c]{$-\hspace{0.07cm}0.106$\\ $\pm$ $0.004$} \\
\hline
\end{tabular}
    \caption{Mandel Q-parameter obtained for the photon number distributions of twisted single photons having different orders generated in parametric down conversion.}
    \label{tab:subP_vortex}
\end{table}

\section{Conclusions}
\label{S:5}

Conventional quantum optics experiments to determine the non-classical statistical behavior of single photon sources involve time-to-digital converters along with associated post-processing. Here, we introduce a simple technique to build the number statistics of heralded single photon source using a digital oscilloscope. The signal and idler in the SPDC output are detected and registered in the oscilloscope. From the recorded data, a time series corresponding to the coincident detections and the corresponding number distribution are obtained. In this way, one can observe the sub-Possonian behavior of the photon number statistics of the source using a set of detectors and an oscilloscope available in almost all the optics laboratories. We also show that in this way we can determine the number statistics of heralded single photons carrying OAM and thus can characterize them for further applications involving OAM of single photons.

%\nocite{*}

%

\end{document}